\documentclass[12pt]{article} 
\usepackage{amsmath, amssymb, amscd}
\usepackage{graphicx}

\begin{document}

\title{Eventually Number-Conserving Cellular Automata}
\author{Nino Boccara\\
Department of Physics\\
University of Illinois at Chicago\\
\texttt{boccara@uic.edu}}

\date{}

\maketitle

\section*{Abstract}   We present a preliminary study of a new class of two-input cellular automata 
called \emph{eventually number-conserving cellular automata} characterized by the property of evolving 
after a finite number of time steps to states whose number of active sites remains constant.  
Eventually number-conserving cellular automata are models of open systems of interacting particles, that is, system of particles interacting with the external world, The particle aspect of eventually number-conserving cellular automata can be emphasized by the motion representation of the cellular automaton evolution rule. This new class of cellular automata contains, as strict subclasses, number-conserving cellular automata, monotone cellular automata, and cellular automata emulating number-conserving ones. Our main objective is to show that they are not what one might naively think they are. 

\section{Introduction} 

Whereas cellular automata have been widely used to model complex systems in which the local character of the evolution rule plays an essential role~\cite{PBVB, G, BGMP, BSS, B}, the theory of cellular automata is still in its infancy and very few exact results---as, for example, the necessary and sufficient condition for a cellular automaton rule, either deterministic or probabilistic, to be number-conserving--- have been obtained~\cite{BF1, BF2, DFR, FG, M, F}  Recently most of the theoretical research carried out into this field is devoted to definition and characterization of new families of rules. In order to extend the family of number-conserving rules a number of papers on monotone rules (the set of number-conserving rules is a subset of monotone rules) have been published~\cite{MBG}. The complexity of problems concerning characterizations of these properties may be illustrated by a recent result~\cite{BBFK} showing that monotony is decidable in dimension one but not in higher dimensions. This paper is devoted to the study of a new class of cellular automata called \emph{eventually number-conserving cellular automata} defined by the property of evolving  after a finite number of time steps to states whose number of active sites remains constant.  The characterization of this new class of cellular automata is challenging since both number-conserving and monotone rules are special cases of eventually number-conserving rules, and it is not, therefore, obvious that eventual number conservation is a decidable property.

A one-dimensional cellular automaton (CA) is a discrete dynamical system, which may be defined as follows. Let $s:\mathbb{Z}\times\mathbb{N}\mapsto\mathcal{Q}$ be a function satisfying the equation
\begin{equation}
s(i,t+1) = f\big(s(i-r_\ell,t),s(i-r_\ell+1,t),\ldots,s(i+r_r,t)\big),
\label{evolution}
\end{equation}
for all $i\in\mathbb{Z}$ and all $t\in\mathbb{N}$ where $\mathbb{Z}$ denotes the set of all integers, $\mathbb{N}$ the set of nonnegative integers, and $\mathcal{Q}$ a finite set of states, usually
equal to $\{0,1,2,\ldots,q-1\}$. $s(i,t)$ represents the \emph{state of site $i$ at time $t$}, and the mapping
$f : {\mathcal{Q}}^{r_\ell+r_r+1}\to\mathcal{Q}$ is the \emph{local CA  evolution rule}. The positive integers $r_\ell$ and $r_r$ are, respectively, the \textit{left} and \textit{right radii} of the rule. In what follows, $f$ will be referred to as an \textit{$n$-input rule}, where $n$ is the number $r_\ell+r_r+1$ of arguments of $f$. Any map $c:\mathbb{Z}_L\mapsto\mathcal{Q}$ is called a \emph{configuration}. Following Wolfram~\cite{W}, to each rule $f$ we assign a \emph{code number} $N(f)$ such that
$$
N(f) = \sum_{(x_1,x_2,\ldots,x_n)\in{\mathcal{Q}}^n}
f(x_1,x_2,\ldots,x_n)q^{q^{n-1}x_1+q^{n-2}x_2+\cdots+q^0x_n}.
$$
Two-state, three-input cellular automata (CAs) are referred to as \emph{elementary} CAs.

In this paper, we will only consider finite (or periodic) CAs, and replace the set $\mathbb{Z}$ by the set 
${\mathbb{Z}}_L$ of integers modulo $L$. Any element of the set ${\mathcal{Q}}^L$ will be called a 
\emph{cyclic configuration of length $L$}. 

\textit{A one-dimensional $q$-state $n$-input 
CA rule $f$ is \emph{number-conserving} if, for all cyclic configurations of length
$L\ge n$, it satisfies}
\begin{multline}
f(x_1,x_2,\ldots,x_{n-1},x_n)+f(x_2,x_3,\ldots,x_n,x_{n+1})+\cdots\\
+f(x_L,x_1\ldots,x_{n-2},x_{n-1})=x_1+x_2+\cdots+x_L.
\label{NCdef}
\end{multline}

It can be shown~\cite{BF2} that: 

\textit{A one-dimensional $q$-state $n$-input 
CA rule $f$ is number-conserving if, and only if, for all
$(x_1,x_2,\ldots,x_n)\in{\mathcal{Q}}^n$, it satisfies}
\begin{align}
f(x_1,x_2,\ldots,x_n) = x_1 + \sum_{k=1}^{n-1}\big(
&f(\underbrace{0,0,\ldots,0}_k,x_2,x_3,\ldots,x_{n-k+1})\notag\\
-&f(\underbrace{0,0,\ldots,0}_k,x_1,x_2,\ldots,x_{n-k})\big),
\label{NScond}
\end{align}

The purpose of this paper is to study \textit{eventually number-conserving CA rules}.
This new class of CA rules, which is an extension of the class of 
number-conserving CA rules, is defined as follows:

\textit{A one-dimensional $q$-state $n$-input CA rule $f$
is \emph{eventually number-con\-ser\-ving} (ENC) if, after a \emph{finite} number of iterations of
rule $f$, it satisfies condition~(\ref{NCdef}) for all cyclic configurations of length $L\ge n$.}

While number-conserving CAs may be viewed as models of isolated systems of interacting particles in which processes of annihilation or creation of particles are forbidden, ENC CAs are models of open systems of interacting particles exchanging particles with the external world such that at after a finite transient time the numbers of particles moving from and to the system exactly counterbalance each other.  Numerical simulations show that the constant number of active sites in configurations of the limit set of an ENC CA  depend not only upon the initial number of active sites but also on the detailed structure of the initial configuration. 

In the two following sections we consider classes of CAs that one might think coincide with ENC CAs.  In the last section the \emph{motion representation} of CA rules is introduced. More explicitly than the usual rule table, it shows how to represent  particle motion either inside the system or between the system and the external world, stressing the fact that ENC CAs can be viewed as systems of interacting particles.

\section{Monotone CA rules}

\textit{A one-dimensional $q$-state $n$-input CA rule $f$ is \emph{number-nondecreasing} if, for all cyclic configurations of length $L\ge n$, it satisfies
\begin{multline}
f(x_1,x_2,\ldots,x_{n-1},x_n)+f(x_2,x_3,\ldots,x_n,x_{n+1})+\cdots\\
+f(x_L,x_1\ldots,x_{n-2},x_{n-1}) \geq x_1+x_2+\cdots+x_L.
\label{NDR}
\end{multline}}

The $q$-state $n$-input CA rule $C(f)$,  defined by
$$
C(f)(x_1,x_2,\ldots,x_n) = f(q-1-x_1, q-1-x_2,\ldots,q-1-x_n),
$$
is called the \emph{conjugate} of rule $f$. This definition implies that the conjugate of  a number-nondecreasing CA rule $f$ is \emph{number-nonincreasing}, that is, for all cyclic configurations of length
$L\ge n$, it satisfies
\begin{multline}
f(x_1,x_2,\ldots,x_{n-1},x_n)+f(x_2,x_3,\ldots,x_n,x_{n+1})+\cdots\\
+f(x_L,x_1\ldots,x_{n-2},x_{n-1}) \leq x_1+x_2+\cdots+x_L.
\label{NIR}
\end{multline}

A CA rule  is \emph{monotone} if it is either  number-nondecreasing or number-nonincreasing. Some properties of monotone CAs have been discussed in~\cite{MBG}.

If a CA rule is \emph{number-nondecreasing} and \emph{number-nonincreasing},
it is  \emph{number-conserving}. The class of number-conserving rules is therefore a subclass of the class of monotone rules.

It is clear that all monotone rules are ENC, but have we a theorem stating a necessary and sufficient condition for a CA rule $f$ to be monotone?  The answer is  ``yes.''  It can be shown~\cite{BF3}, that\footnote{Another, less practical, necessary and sufficient condition can be found in~\cite{MBG}.}

 \textit{A one-dimensional $q$-state $n$-input CA rule $f$ is \emph{number-nondecreasing} (resp. \emph{number-nonincreasing})  if, for all cyclic configurations of lengths $2n-2$ \emph{and} $2n-1$ it satisfies relation~(\ref{NDR}) (resp. (\ref{NIR}))}.

\noindent\textbf{Remark.} The word \emph{and} is essential since the CA rules that verify (\ref{NDR}) or (\ref{NIR}) for all cyclic configurations of either length $2n-2$ or $2n-1$ are not necessarily monotone. To be monotone, say nondecreasing, a rule has to verify relation~(\ref{NDR}) for all configurations of lengths  $2n-2$ and $2n-1$. This is, in particular, the case for three-state two-input CA rules. The sets of rules that satisfy relation~(\ref{NDR}) for all cyclic configurations of lengths 2 and 3 have, respectively, 864 and 724 elements while the intersection of these two sets contains only 708 CA rules which are the only monotone nondecreasing rules. 

There exist 87 monotone elementary CA rules. The sets of nonincreasing and nondecreasing elementary CA rules  both contain 46 rules. The intersection of these two sets consists of the five number-conserving elementary CA rules.\footnote{See the appendix for a complete list.}  

Although all monotone rules are ENC rules, are all ENC rules monotone? The answer is ``no.''
In other words, the class of monotone rules is \emph{strictly} included in the class of ENC rules.
For example, elementary CA rules 99, 173, and 229  are ENC but not monotone. They are defined, respectively, by
\begin{alignat*}{4}
f_{99}(0,0,0) & = 1, &\quad f_{99}(0,0,1) & = 1, &\quad f_{99}(0,1,0) & = 0,
&\quad f_{99}(0,1,1) & = 0,\\
f_{99}(1,0,0) & = 0, &\quad f_{99}(1,0,1) & = 1, &\quad f_{99}(1,1,0) & = 1,
&\quad f_{99}(1,1,1) & = 0,
\end{alignat*}
\begin{alignat*}{4}
f_{173}(0,0,0) & = 1, &\quad f_{173}(0,0,1) & = 0, &\quad f_{173}(0,1,0) & = 1,
&\quad f_{173}(0,1,1) & = 1,\\
f_{173}(1,0,0) & = 0, &\quad f_{173}(1,0,1) & = 1, &\quad f_{173}(1,1,0) & = 0,
&\quad f_{173}(1,1,1) & = 1,
\end{alignat*}
and
\begin{alignat*}{4}
f_{229}(0,0,0) & = 1, &\quad f_{229}(0,0,1) & = 0, &\quad f_{229}(0,1,0) & = 1,
&\quad f_{229}(0,1,1) & = 0,\\
f_{229}(1,0,0) & = 0, &\quad f_{229}(1,0,1) & = 1, &\quad f_{229}(1,1,0) & = 1,
&\quad f_{229}(1,1,1) & = 1.
\end{alignat*}

\section{Emulating number-conserving rules}

\textit{A CA rule $f$ emulates rule $g$ if, for all finite configuration of length $L$ belonging to 
the limit set $\Lambda_f$ of rule $f$, its images by either $f$ or $g$ are equal. That is, for all 
$c\in\Lambda_f $ and all $ i\in\mathbb{Z}_L $, 
$$
g\big(c(i-r_\ell),c(i-r_\ell+1),\ldots,c(i+r_r)\big) = f\big(c(i-r_\ell),c(i-r_\ell+1),\ldots,c(i+r_r)\big).
$$}

It would be tempting to define a ENC CA rule as a CA rule that emulates, at least, one number-conserving rule. Here are two examples.

\begin{figure}[h]
\centering\includegraphics[scale=1.0] {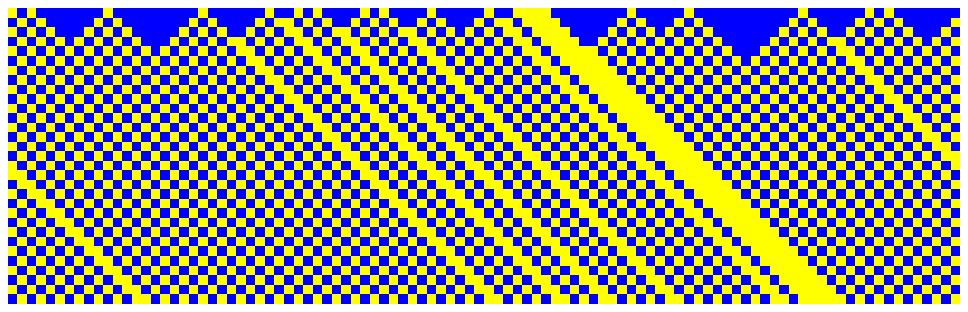}
\centering\includegraphics[scale=1.0] {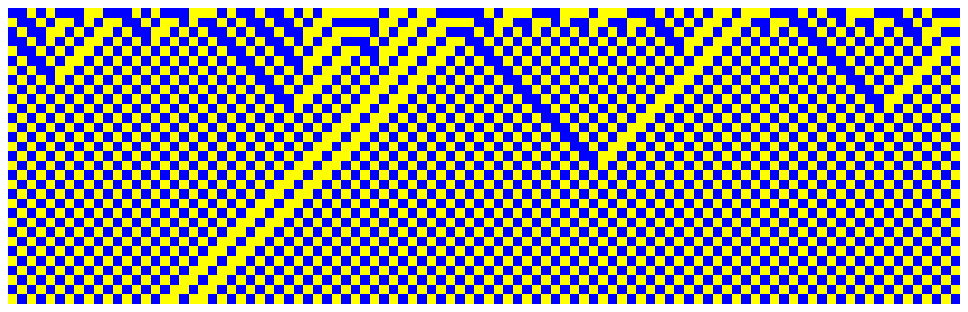}
\caption{\label{fig:pattern176and99} \textit{The spatiotemporal patterns of elementary CA rules 176 (top) and 99 (bottom) show clearly that these rules emulate, respectively, elementary CA rules 184 and 240, and elementary rules 226.}}
\end{figure}

\noindent\textbf{Example 1.} Elementary CA rule 176, defined by
\begin{alignat*}{4}
f_{176}(0,0,0) & = 0, &\quad f_{176}(0,0,1) & = 0, &\quad f_{176}(0,1,0) & = 0,
&\quad f_{176}(0,1,1) & = 0,\\
f_{176}(1,0,0) & = 1, &\quad f_{176}(1,0,1) & = 1, &\quad f_{176}(1,1,0) & = 0,
&\quad f_{176}(1,1,1) & = 1,
\end{alignat*}
is number-nonincreasing. One can readily verify that it emulates number-conserving elementary CA rules 184 and 240 (see top figure~\ref{fig:pattern176and99}). Configurations belonging to its limit set 
consist of isolated 1s separated by sequences of 0s whose lengths depend upon the initial configuration.

\noindent\textbf{Example 2.} Elementary CA rule 99 defined above, which is not monotone, emulates  elementary CA rule 226. Configurations of its limit set consist of alternating sequences of 0s and 1s with, depending upon the initial configuration, either a few pairs of 0s, separating two successive 1s, or a few pairs of 1s, separating two successive 0s (see bottom figure~\ref{fig:pattern176and99}).

In Appendix 2 we show how to determine ENC CA rules emulating number-conserving rules.

\begin{figure}[h]
\centering\includegraphics[scale=1.0] {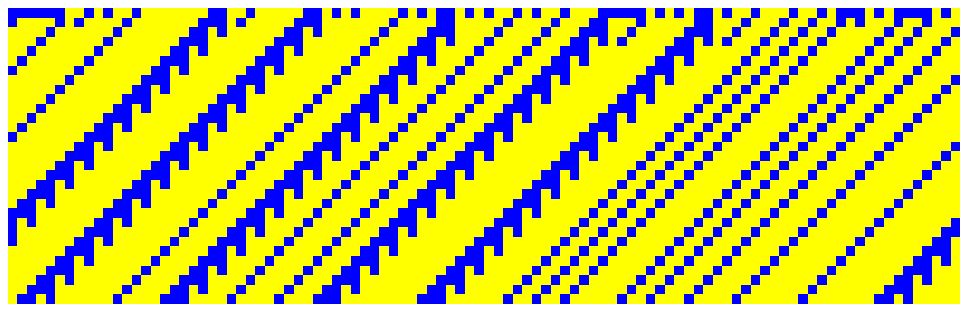}
\centering\includegraphics[scale=1.0] {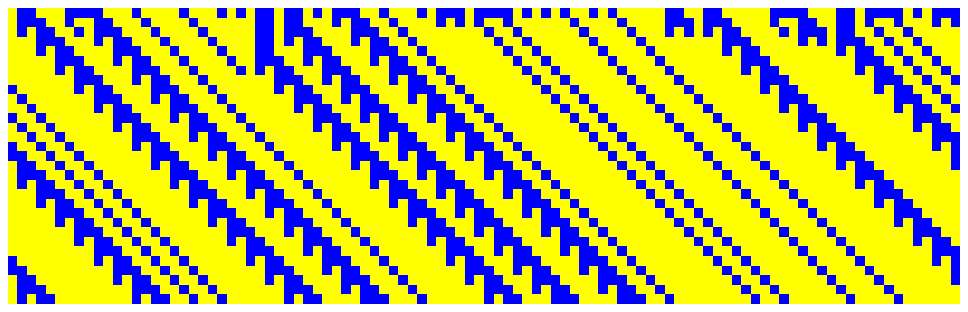}
\caption{\label{fig:patterns74and88} \textit{Spatiotemporal patterns of elementary CA 
rules 74 (top) and 88 (bottom). Both rules---one being the reflected rule of the other---are 
not monotone and do not emulate number-conservative rules. Initial configurations are random.}}
\end{figure}

Are there ENC CA rules that do not emulate a number-conserving rule? Surprisingly, 
the answer is  ``yes,'' and defining ENC CA rules as rules, which after a finite number
of iterations emulate number-conserving rules would be inadequate. As for monotone rules, the set of rules that emulate a number-conserving rules is \emph{strictly} included in the set of ENC rules. For example, elementary CA rules 74 and 88, obtained by reflection\footnote{The $q$-state $n$-input CA rule $R(f)$, defined by
$$
R(f)(x_1,x_2,\ldots,x_n) = f(x_n,x_{n-1},\ldots,x_1),
$$
is called the \emph{reflected} of rule $f$.} of rule 74, are examples of such rules. 
They are defined by
\begin{alignat*}{4}
f_{74}(0,0,0) & = 0, &\quad f_{74}(0,0,1) & = 1, &\quad f_{74}(0,1,0) & = 0,
&\quad f_{74}(0,1,1) & = 1,\\
f_{74}(1,0,0) & = 0, &\quad f_{74}(1,0,1) & = 0, &\quad f_{74}(1,1,0) & = 1,
&\quad f_{74}(1,1,1) & = 0,
\end{alignat*}
and
\begin{alignat*}{4}
f_{88}(0,0,0) & = 0, &\quad f_{88}(0,0,1) & = 0, &\quad f_{88}(0,1,0) & = 0,
&\quad f_{88}(0,1,1) & = 1,\\
f_{88}(1,0,0) & = 1, &\quad f_{88}(1,0,1) & = 0, &\quad f_{88}(1,1,0) & = 1,
&\quad f_{88}(1,1,1) & = 0,
\end{alignat*}
They are not monotone and do not emulate number-conserving rules. Actually, configurations belonging to their limit sets contains all the 8 different triplets. Both spatiotemporal patterns exhibit the propagation 
in opposite directions of similar structures (see figure~\ref{fig:patterns74and88}). Rules 173 and 229, defined above, and that are respectively conjugate of rules 74 and 88, do not emulate number-conserving rules and have spatiotemporal patterns that, like rules 74 and 88, exhibit the propagation in opposite directions of similar structures.

Since there exist ENC rules that are not monotone and do not emulate any number-conserving one, the union of the set of monotone rules and the set of rules emulating number-conserving rules is \emph{strictly} included in the set of ENC rules.

\section{Motion representation}
The particle dynamics is not clearly exhibited by the CA rule table. A simpler and more visual picture of the evolution rule can be given by its \emph{motion representation}.  This representation, which has been first introduced for number-conserving rules~\cite{BF1}, was  defined as follows. List all the neighborhoods of an occupied site represented by its site value $s\in\mathcal{Q}$. Then, for each neighborhood, indicate the displacements of the $s$ particles by  arrow(s) joining the site where the particle(s) is (are) initially located to its (their) final position(s). A number above the arrow indicates how many particles are moving to the final position. To simplify the representation only neighborhoods for which, at least one particle is moving. are listed. Here are two examples. The motion representation of the two-state three-input rule 184, which represents the simplest car moving rule on a one-lane highway, is
$$
\overset{\overset{1}{\curvearrowright}}{10}.
$$
Since a particle located on the left of an occupied site cannot move to the right (which the direction of motion) we do not mention it.  In a less compact notation we could complete the above representation with 
$$
\overset{\circlearrowleft}{1}1.
$$
The motion representation of the three-state three-input rule 6171534259461 is
$$
\overset{\overset{1}{\curvearrowright}}{10}\quad\overset{\overset{1}{\curvearrowright}}{11}\quad
\overset{\overset{2}{\curvearrowright}}{20}\quad\overset{\overset{1}{\curvearrowright}}{21}
$$
In both examples particles move only to the right.  It is, therefore, not necessary to indicate the state of the left neighboring site of the particle(s).

It is rather straightforward to extend this representation to nonconservative rules. The only difference with number-conserving rules is to indicate the possibility of creations and annihilations of particles. This is done by adding either a $+$ or a $-$ sign above the site state followed by a positive number $n$ to represent, respectively, creation and annihilation of $n$ particles at that site.  Although the number of particles is not conserved, we shall keep the name \emph{motion representation}, since creations and annihilations of particles may be viewed as particles moving between the system and the external world.
Here are two examples illustrating that ENC CAs rules can be viewed as evolving systems of particles and not just finite sequences of nonnegative integers obeying abstract transformation rules.

\noindent\textbf{Example 1.} The motion of particles in a system evolving according to CA rule 176 is represented by\footnote{A systematic determination of motion representations is given in~\cite{BF3}.}
$$
\overset{\overset{1}{\curvearrowright}}{10}\; \bullet, \quad 0{\overset{-1}1}\bullet,
$$
where the symbol $\bullet$ may represent either a 0 or a 1.  The second term shows that only isolated 1s can survive, as illustrated in top figure~\ref{fig:pattern176and99}. In addition, when no more particles can be annihilated, that is, when all particles have empty neighboring sites, the evolution rule of the system coincides with CA rules 184 and 240, in agreement with the fact that rule 176 emulates these two rules.

\noindent\textbf{Example 2.} The evolution of a system of particles evolving according to CA rule 74 is represented by
$$
0\overset{\overset{1}{\curvearrowleft}}{01}, \quad 0{\overset{-1}1}0,\quad  1{\overset{-1}1}1.
$$
The first term shows that isolated 1s move to the left (see top figure~\ref{fig:patterns74and88}), furthermore, starting from a triplet of 1s in a sea of 0s, the combination of these three terms generates t a composite particle of length 6 oscillating between two states: $001110$ and $011010$ propagating to the left, or the two-row composite particle
\begin{align*}
& 0 0 1 1 1 0\\
& 0 1 1 0 1 0
\end{align*}
which, then propagates to the left. Note that the motion representation lists only neighborhoods of moving particles, where ``moving'' stands for motion to or from the external world.

\section{Conclusion}

We have introduced a new class of CAs called  \textit{eventually number-conserving} (ENC) CAs  adopting the following definition:  \textit{A one-dimensional $q$-state $n$-input CA rule $f$ is \emph{eventually number-conserving} if, after a \emph{finite} number of iterations, it satisfies the number-conserving condition~(\ref{NCdef}) for all cyclic configurations of length $L\ge n$}. ENC CAs are models of open systems of interacting particles exchanging particles with the external world but that, after a finite transient time, the numbers of particles moving from and to the system exactly counterbalance each other. The interpretation in terms of particles is emphasized by the motion representation of the CA evolution rule. We have shown that the set of ENC CA rules \emph{strictly} includes the set of number-conserving rules, the set of monotone rules, and the set of rules emulating a number-conserving rule. The set of ENC rules also strictly includes the union of all these sets showing that the set of ENC rules  actually contains rules which do not belong to any of these sets. This challenging result suggests that the characterization of this new apparently simple class is probably not a trivial problem. 

\section*{Appendix 1}
The code numbers of monotone nonincreasing  and nondecreasing elementary CA rules are respectively:
\begin{align*}
&\{0, 2, 4, 8, 10, 12, 16, 24, 32, 34, 40, 42, 48, 56, 64, 66, 68, 72, 76, 80, 
96, 98, 112,\\ 
&128,130, 132, 136, 138, 140, 144, 152, 160, 162, 168, 170, 176, 
184, 192, 194,\\ 
&196, 200, 204, 208, 224, 226, 240\},
\end{align*}
and
\begin{align*}
&\{170, 171, 174, 175, 184, 185, 186, 187, 188, 189, 190, 191, 204, 205, 206, 207, \\
&220,221, 222, 223, 226, 227, 230, 231, 234, 235, 236, 237, 238, 239, 
240, 241,\\ 
&242, 243,244, 245, 246, 247, 248, 249, 250, 251, 252, 253, 254, 255\}.
\end{align*}
There are five number-conserving elementary CA rules given by the intersection of these two 
sets. Their code numbers are:  $\{170, 184, 204, 226, 240\}$.

\section*{Appendix 2}
In this appendix we show how to find examples of ENC CA rules emulating given number-conserving CA rules.

\begin{figure}[h]
\centering\includegraphics[scale=1.0] {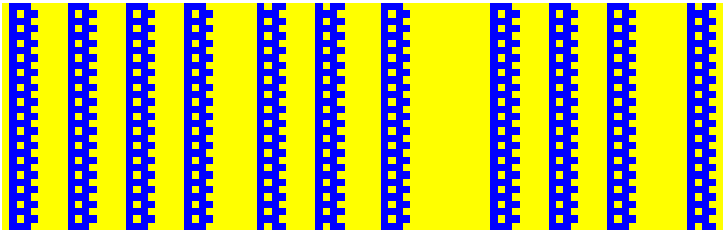}
\caption{\label{fig:limitSet50358} \textit{Spatiotemporal pattern showing the limit set of 4-input rule 50358, where left and right radii have been chosen equal, respectively, to 3 and 0.}}
\end{figure}

Four-input rule 50358 is number-conserving. It is defined by
\begin{alignat*}{4}
f_{50358}(0,0,0,0) & = 0, &\quad f_{50358}(0,0,0,1) & = 1, &\quad f_{50358}(0,0,1,0) & = 1,\\
 f_{50358}(0,0,1,1) & = 0, &\quad f_{50358}(0,1,0,0) & = 1, &\quad f_{50358}(0,1,0,1) & = 1, \\
f_{50358}(0,1,1,0) & = 0, &\quad f_{50358}(0,1,1,1) & = 1, &\quad f_{50358}(1,0,0,0) & = 0, \\
f_{50358}(1,0,0,1) & = 0, &\quad f_{50358}(1,0,1,0) & = 1, &\quad f_{50358}(1,0,1,1) & = 0,\\
f_{50358}(1,1,0,0) & = 0, &\quad f_{50358}(1,1,0,1) & = 0, &\quad f_{50358}(1,1,1,0) & = 1,\\
& &\quad f_{50358}(1,1,1,1) & = 1.
\end{alignat*}
As shown in figure~\ref{fig:limitSet50358}, its limit set, when the left and right radii are respectively equal to 3 and 0, can be viewed as 2-row tiles of one of the following types:
\begin{center}
1110000 \hspace{1cm} 1011000\\
1011000 \hspace{1cm} 1110000\\
\end{center}
concatenated with any number of 2-row tiles of type:
\begin{center}
0 \\
0 \\
\end{center}
Among the 16 different quadruplets of 0s and 1s, the limit set of rule 50358 contains only the following ones:
\begin{alignat*}{4}
&(0, 0, 0, 0),\  &(0, 0, 0, 1),\  &(0, 0, 1, 0),\  &(0, 0, 1, 1), \ &( 0, 1, 0, 1),\\
&(0, 1, 1, 0),\  &(0, 1, 1, 1),\  &(1, 0, 0, 0),\  &(1, 0, 1, 1),\  &(1, 1, 0, 0),\\ 
&&&(1, 1, 1, 0).
\end{alignat*}
Among all the four-input rules having the same images as rule 50358 for all these 11 quadruplets one can verify that only rules 16566, 49334, and 50870 are ENC and have the same limit set as rule 50358. 
These three rules are not monotone.


\begin{thebibliography}{0000}

\bibitem{PBVB} P. Manneville, N. Boccara, G. Vichniac and R. Bidaux
eds., \textit{Cellular Automata and Modeling of Complex Systems,
Proc. of a Workshop, Les Houches} (Heidelberg: Springer-Verlag,
1989).

\bibitem{G} H. Gutowitz  ed., \textit{Cellular Automata: Theory and
Experiments, Proc. of a Workshop, Los Alamos} (Amsterdam:
North-Holland, 1990).

\bibitem{BGMP} N. Boccara, E. Goles, S. Mart\'{\i}nez, and P.
Picco eds., \textit{Cellular Automata and Cooperative Phenomena,
Proc. of a Workshop, Les Houches} (Dordrecht: Kluwer, 1993).

\bibitem{BSS} S. Bandini, R. Serra, and F. Suggi Liverani eds.,
\textit{Cellular Automata: Research Towards Industry, Proceedings of
the Third Conference on Cellular Automata for Research and
Industry} (Heidelberg: Springer-Verlag, 1998).

\bibitem{B} N. Boccara, \textit{Modeling Complex Systems} (New York: 
Springer-Verlag, 2004).

\bibitem{BF1} N. Boccara and H. Fuk\'s, \textit{Cellular Automaton  Rules Conserving the Number of Active Sites} , Journal of Physics A: Mathematical and General {\bf 31} 6007--6018 (1998).

\bibitem{BF2} N. Boccara and H. Fuk\'s, \textit{Number-Conserving 
Cellular Automaton Rules}, Fundamenta Informaticae {\bf 52} 1--13 (2002).

\bibitem{DFR}  B. Durand, E. Formenti, and Z. R\'oka, 
\textit{Number-Conserving Cellular Automata I: Decidability},
Theoretical Computer Science {\bf 299} 523--535 (2003).

\bibitem{FG} E. Formenti and A. Grange, \textit{Number-Conserving 
Cellular Automata II: Dynamics},
Theoretical Computer Science {\bf 304} 269--290 (2003).

\bibitem{M} A. Moreira, \textit{Universality and Decidability of 
Number-Conserving Cellular Automata}, Theoretical Computer 
Science {\bf 292} 711-721 (2003).

\bibitem{F} H. Fuk\'s, \textit{Probabilistic cellular automata with conserved quantities}, 
Nonlinearity {\bf 17} 159--173 (2004).

\bibitem{MBG} A. Moreira, N. Boccara, and E. Goles, \textit{On
Conservative and Monotone One-Dimensional Cellular Automata and 
Their Particle Representation}, Theoretical Computer Science 
{\bf 325} 285--316 (2004).

\bibitem{BBFK} V. Bernardi. B, Durand. E, Formenti. and J, Kari. ;textit{A
new dimension-sensitive property for cellular automata}, 
Proceedings of MFCS 2004,
Lecture Notes in Computer Science, (New York: Springer-Verlag 2004).

\bibitem{W} S. Wolfram, \textit{Cellular Automata and Complexity: Collected
Papers}, (Reading: Addison-Wesley, 1994).

\bibitem{BF3} N. Boccara and H. Fuk\'s, \textit{One-dimensional monotone cellular 
automata}, \texttt{nlin.CG/0501043}.


\end{thebibliography}
\end{document}